\documentclass[10pt,journal,compsoc]{IEEEtran}
\pdfoutput=1

\ifCLASSOPTIONcompsoc
  \usepackage[nocompress]{cite}
\else
  \usepackage{cite}
\fi

\usepackage{caption,subcaption}
\usepackage{listings}
\AtBeginDocument{\DeclareCaptionSubType{lstlisting}}
\usepackage{bp-listings}
\newcommand{\lstin}[2][]{\lstinline[style=,language=,basicstyle=\ttfamily,#1]|#2|}

\PassOptionsToPackage{hyphens}{url}\usepackage[colorlinks]{hyperref}
\usepackage{adjustbox}
\usepackage{amsmath}
\usepackage{amsthm}
\usepackage{amssymb}
\usepackage{orcidlink}
\usepackage{graphicx}
\usepackage{booktabs}
\usepackage{tabularx}
\usepackage{array}
\newcolumntype{P}[1]{>{\centering\arraybackslash}p{#1}}
\usepackage{multirow}
\usepackage{tikz}
\usetikzlibrary{arrows,automata,positioning,shapes,patterns.meta}
\usepackage{enumitem}

\newtheorem{definition}{Definition}

\newtheorem{example}{Example}
\usepackage{xcolor, soul}

 \usepackage{booktabs}

\usepackage{pgfplots}

\usepackage{marginnote}

\pgfplotsset{width=10cm, compat=1.9}

\begin{document}

\title{Generalized Coverage Criteria for\\Combinatorial Sequence Testing}

\author{Achiya~Elyasaf~\orcidlink{0000-0002-4009-5353},
        Eitan~Farchi~\orcidlink{0000-0002-3021-1488},
        Oded~Margalit~\orcidlink{0000-0002-2026-2601},
        Gera~Weiss~\orcidlink{0000-0002-5832-8768},
        and~Yeshayahu~Weiss~\orcidlink{0000-0002-8183-5282},
\IEEEcompsocitemizethanks{%
\IEEEcompsocthanksitem A. Elyasaf is with the Software and Information Systems Engineering Department, Ben-Gurion University of the Negev, Israel.\protect\\
E-mail: \texttt{achiya@bgu.ac.il}
\IEEEcompsocthanksitem O. Margalit, G. Weiss, and Y. Weiss, are with the Computer Science Department, Ben-Gurion University of the Negev, Israel.\protect\\
E-mail: \texttt{odedm@post.bgu.ac.il}, \texttt{geraw@cs.bgu.ac.il}, and \texttt{weissye@post.bgu.ac.il}
\IEEEcompsocthanksitem E. Farchi is with IBM Haifa Research Lab.\protect\\
E-mail: \texttt{farchi@il.ibm.com}}
}

\markboth{IEEE Transactions on Software Engineering}%
{Elyasaf \MakeLowercase{\textit{et al.}}: Generalized Coverage Criteria for Combinatorial Sequence Testing}
%


\IEEEtitleabstractindextext{%

\begin{abstract}
We present a new model-based approach for testing systems that use sequences of actions and assertions as test vectors. Our solution includes a method for quantifying testing quality, a tool for generating high-quality test suites based on the coverage criteria we propose, and a framework for assessing risks.
For testing quality, we propose a method that specifies generalized coverage criteria over sequences of actions, which extends previous approaches. Our publicly available tool demonstrates how to extract effective test suites from test plans based on these criteria. We also present a Bayesian approach for measuring the probabilities of bugs or risks, and show how this quantification can help achieve an informed balance between exploitation and exploration in testing.
Finally, we provide an empirical evaluation demonstrating the effectiveness of our tool in finding bugs, assessing risks, and achieving coverage.
\end{abstract}

}

\maketitle

\IEEEdisplaynontitleabstractindextext

%
\IEEEpeerreviewmaketitle

\newcounter{missing}
\newcommand{\missing}[1]{\refstepcounter{missing}\marginpar{\textsuperscript{\arabic{missing}}#1}}

\IEEEraisesectionheading{\section{Introduction}\label{sec:intro}}
\IEEEPARstart{T}{esting} faces one of its biggest challenges is knowing when to stop. How many tests are necessary? Should we aim to test every possible input or focus on inputs that are more likely to cause errors? Is it better to allocate our resources uniformly throughout the application or invest our efforts in the most critical paths?

To address these problems in black-box testing, several techniques have been proposed for providing a set of input vectors for testing the system. These techniques define coverage criteria on the possible input vectors, thus dramatically reducing the nearly infinite possibilities. These techniques include, for example, decision table testing, domain analysis, all-pairs testing, boundary-value analysis, and more~\cite{forgacs2019practical}.

The aforementioned techniques test the possible system behaviors by analyzing the effect of the input vector on the system behavior and then clustering the input-vector space. Another type of technique clusters the system-behavior space, either by clustering the behaviors manually (e.g., by use cases or user stories) or systematically. Kuhn and Higdon\footnote{We attribute this contribution to Khun and Higdon since the book ``Practical Combinatorial Testing" by Kuhn, Kacker, and Lei~\cite{kuhn2010practical} references an earlier work from 2009 by D.R. Kuhn and J.M. Higdon entitled ``Testing Event Sequences". However, we could not obtain a copy of this paper.} proposed a coverage criterion that is based on sequences of events of a system, called \textit{$t$-way sequence coverage}, where sequences of any $t$ events cluster the system behavior~\cite{kuhn2010practical, maximoff2010method, duan2019approach}.

In this paper, we continue and extend Kuhn and Higdon's approach for a direct clustering of the system behavior best defined as a sequence of events.
More generally, we propose to address the above testing challenge by tackling the following issues:
(1) how to cluster the space of event sequences that need to be covered; (2) how to cover this space with a finite, relatively small number of tests; and (3) how to methodically explore and exploit knowledge from previous tests to optimally reduce bug risks. 

Our answer to the first challenge is to define generalized coverage criteria for sequences of events. With examples from various domains, we demonstrate how our generalizations can naturally express useful coverage criteria that extend Kuhn and Higdon's vision by allowing more types of applications. We also show how the proposed extension enables the inclusion of standard combinatorial-test-design (CTD) methods and Kuhn and Higdon's criterion under a unified formal vocabulary~\cite{CTD}. Our approach to specifying coverage criteria is automata-based. Each automaton in our framework represents a set of tests considered equivalent by testers. For example, two tests in the same set are likely to both fail or both pass, presumably because the test designer expects that the probability that the system under test handles them differently is low. We present various examples from different systems to demonstrate the naturalness of this approach and show that testers can construct coverage requirements to suit different domains irrespective of a specific test model.

Our answer to the second challenge is a publicly available tool for test-suites generation that demonstrates how our generalized criteria can be used in practice. Our sequence coverage paradigm includes the definition of a set of automata. Testers can provide their test coverage criteria by providing a ranking function that counts the number of automata (in their coverage criteria) that pass at least one of the tests in a given test suite. A genetic-algorithms (GA) based technique is then applied to produce high-ranking test suites. 

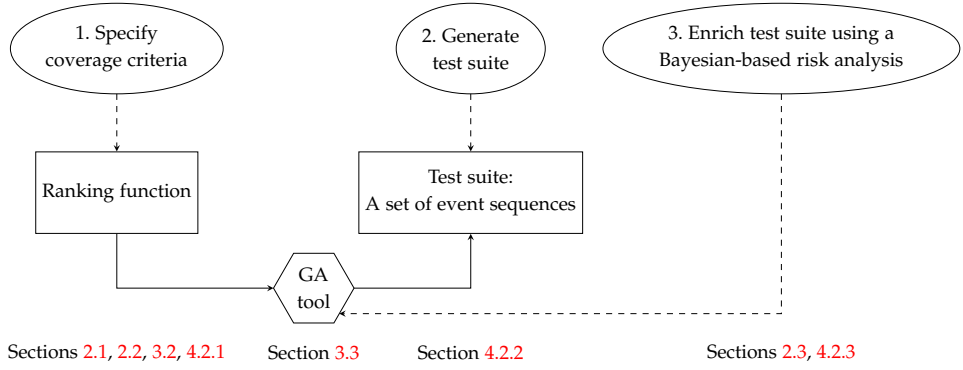
\begin{figure*}
\centering
\begin{adjustbox}{width=0.7\linewidth}
\begin{tikzpicture}[
ellipsenode/.style={ellipse,draw,align=center,inner sep=2,outer sep=0,node distance=0.3cm and 0.2cm,minimum height=10ex,execute at begin node=\setlength{\baselineskip}{3ex}},
squarednode/.style={rectangle,draw,align=center,minimum height=9ex,node distance=0.7cm and 0.3cm,execute at begin node=\setlength{\baselineskip}{3ex}},
polygonnode/.style={regular polygon,regular polygon sides=6,draw,align=center,inner sep=2,outer sep=0,node distance=0.3cm and 0.2cm,minimum height=8ex,execute at begin node=\setlength{\baselineskip}{3ex}},
textnode/.style={align=center}
]
\node[ellipsenode] (scc) {1. Specify\\coverage criteria};
\node[ellipsenode] (gts) [right=3cm of scc] {2. Generate\\test suite};
\node[ellipsenode] (bay) [right=1cm of gts] {3. Enrich test suite using a\\Bayesian-based risk analysis};

\node[squarednode] (rf) [below=1cm of scc] {Ranking function};
\node[squarednode] (ts) at (gts |- rf) {Test suite:\\A set of event sequences};
\node[polygonnode] (poc) [below right=0.5cm and 1.56cm of rf] {GA\\tool};

\node[textnode] (trf) [below=1.8cm of rf] {Sections~\ref{sec:approach:generalized},~\ref{sec:approach:examples},~\ref{sec:tool:coverage},~\ref{sec:eval:criteria}};
\node[textnode] (tpoc) at (poc|-trf) {Section~\ref{sec:tool:generation}};
\node[textnode] (tts) at (ts|-trf) {Section~\ref{sec:eval:generation}};
\node[textnode] (tbay) at (bay|-trf) {Sections~\ref{sec:approach:bayesian},~\ref{sec:eval:bayesian}};

\draw[dashed,-stealth] (scc) -- (rf);
\draw[dashed,-stealth] (gts) -- (ts);
\draw[dashed,-stealth] (bay) |- (poc.south east);
\draw[-stealth] (rf) |- (poc);
\draw[-stealth] (poc) -| (ts);
\end{tikzpicture}
\end{adjustbox}
\caption{A schema of our approach. It consists of three phases; each has an activity (oval) that generates an artifact (box). The test suite is generated using the GA-based, test-suite-generation tool (hexagon) that takes the system-behavior model and the ranking function and returns a set of sequences of events that achieve a high coverage rank. Below the tool and each phase are their relevant section numbers.}
\label{fig:methodology}
\end{figure*}

Our answer to the third challenge of balancing exploitation and exploration is as follows. We acknowledge that the specification of the coverage criteria is not dichotomic, i.e., two tests may belong to the language of the same automaton, but one may exhibit a problem while the other does not.  We thus apply a Bayesian approach. The Bayesian framing of the testing process yields a mathematical formula for balancing exploration and exploitation of the knowledge obtained by tests.  Specifically, we consider each class of equivalent tests as a Bernoulli random variable with a certain probability of hitting a bug. We then show how to measure and use these probabilities to maximize the likelihood of finding new bugs.

We propose a methodology with three activities (depicted in \autoref{fig:methodology}) and demonstrate it through our test-suite-generation tool. Given a model of the system behavior (specified as, e.g., automata), it starts by specifying the ranking function. The two artifacts are then used to generate a test suite (i.e., a set of event sequences) that achieves a high coverage ranking. Finally, a Bayesian-based approach is applied to balance the need to exploit knowledge from previous tests and explore new areas. We envision practitioners defining their domain-specific coverage criteria and selecting predefined generalized criteria from a library. The selection can be based on, for example, the type of implementation at hand, the project phase, the code libraries in use, etc.

The paper is organized as follows. We begin with a theoretical presentation of our approach in~\autoref{sec:approach}. \autoref{sec:tool}~presents our publicly available tool and demonstrates our approach and methodology through this tool. This section also provides a qualitative evaluation of our approach, demonstrating its applicability for testing. We conclude with a quantitative evaluation of our approach in~\autoref{sec:evaluation}.

\section{A Model-Based Approach to Sequence Testing}
\label{sec:approach}
In this section, we present the conceptual approach proposed in this paper. We begin with our new coverage criteria and continue with a formulation of a Bayesian approach to risk assessment. Note that these contributions are independent and can be used separately. To simplify the reading of this section, the main mathematical notations we use are summarized in \autoref{tab:notations}.

\subsection{Generalized Coverage Criteria}
\label{sec:approach:generalized}
We propose a new type of coverage criteria that generalizes the $t$-way combinatorial sequence coverage criterion~\cite{kuhn2010practical} 
and the classical $t$-way coverage. The generalized framework gives testers tools for infusing domain knowledge into the test design. For example, based on previous experience with the system or an understanding of its implementation, a tester may want to focus on testing some race conditions on the access of one shared resource only when another shared resource is held. We define our generalized coverage notion and then motivate it with examples.

\setlength\extrarowheight{2pt}
\begin{table}[tbp]
\caption{A summary of the main mathematical notations.}
\label{tab:notations}
\centering
\begin{tabularx}{\linewidth}{c|X}
\toprule
Symbol                      & Explanation                                               \\ \midrule\midrule
$\Sigma$                    & Alphabet of test actions like `ClickLoginButton'.      \\ 
$a$                         & Length of the test sequences (a number or $*$).           \\ 
$P \subseteq \Sigma^a$      & A regular language of possible test sequences.            \\ 
$I$                         & An index set for naming the coverage sets.                \\ 
$C(i)$, for $i \in I$       & A set of tests that cover some aspect.                    \\ 
$C=\{C(i)\}_{i\in I}$       & A coverage criterion specifying the aspects.              \\
\bottomrule
\end{tabularx}
\end{table}
\setlength\extrarowheight{0pt}


We consider the test model as a set of all possible tests. Each is a sequence of actions that may be applied to the system under test (SUT) using an external interface available for testing. More formally, it is an abstract regular set $P \subseteq \Sigma^a$ where $\Sigma$ is an alphabet that represents the possible actions, and $a \in \mathbb{N} \cup \{*\}$, meaning that each test may be of any length. 

Once $P$ is defined, we would ideally have liked to execute each $t \in P$ against the SUT, but this may be impossible to realize, as $P$ may be huge or even infinite. We, therefore, propose to use coverage criteria to overcome that. We define coverage criteria $C$ using an indexed set of languages $C=\{C(i)\}_{i\in I}, C(i) \subseteq \Sigma^a$. The following definition captures what constitutes covering $P$ using $C$.  

\begin{definition}[coverage]
A set of tests $S \subseteq P$ is said to cover a test-model $P \subseteq \Sigma^a$ under the coverage criteria $C=\{C(i)\}_{i\in I}$, $C(i) \subseteq \Sigma^a\,$ if $\,\forall i\in I\colon (C(i) \cap P \neq \emptyset \Rightarrow C(i) \cap S \neq \emptyset)$ and $\bigcup_{i\in I} C(i) = \Sigma^a$.
\label{def:cov_crit}
\end{definition}

As explained above, a tester may want to focus on testing some race conditions on shared resources.  In such a case, $\Sigma^a$ will represent sequences of reads and writes to the shared resources of interest. The $C(i)$s will specify specific races of interest, e.g., two consecutive writes to a shared table by one process, with a read occurring after the first write and before the second by another process.  To describe such requirements more easily by the tester, domain-specific languages (DSL) may be used. 

Coverage is used to analyze and identify missing tests and report on the progress of the test effort. To allow for the latter, we also define the percentage of coverage obtained. 

\begin{definition}[coverage ratio]
Given a finite $I$, a set of tests $S \subseteq P$ for a test-model $P \subseteq \Sigma^a$, and a coverage criterion $C=\{C(i)\}_{i\in I}$, $C(i) \subseteq \Sigma^a $, we define the coverage ratio as:   
$$\Gamma_{C}(S,P)=\frac{|\{ i \in I \colon  C(i) \cap S \neq \emptyset\}|}{|\{ i \in I \colon  C(i) \cap P \neq \emptyset\}|}.$$
\label{def:cov_rank}
\end{definition}

A few comments on the above definitions are in order:
Based on test concerns, the budget allocated for testing, and the expected usage of the system, it is natural in some applications to require that each test falls within some $C(i)$ and that $P \subseteq \bigcup_{i\in I} C(i) \subset \Sigma^a$. To accommodate that, we always implicitly assume that in practice, there is another coverage requirement $\Sigma^a \,\setminus\, \bigcup_{i\in I} C(i)$, even if it has an empty intersection with $P$. 

Note that a coverage requirement may be a single test. Typically, that may represent a happy path scenario the system should meet. For example, if a single user opens a file that exists, the user has read permission and reads one byte successfully from the opened file. 

Another natural expectation is that $\{C(i)\}_{i\in I}$ is a partition. We chose not to enforce that because this requirement may limit the flexibility of testers to define their test concerns freely. Once  $\{C(i)\}_{i\in I}$ is defined, a partition of  $\bigcup_{i\in I} C(i)$ can be easily achieved if desired by considering coverage requirements of the form 
$\tilde{C}(i) = C(i) \,\setminus\, \bigcup_{i'< i} C(i)$, assuming an (arbitrary) order over the index set. As the following example shows, partitioning the coverage requirements is sometimes unnatural. Consider condition coverage, for example, where each condition in the SUT should be evaluated as true by some test and false by another. Given $n$ conditions in the software, we will get $2n$ coverage requirements, namely, that the first condition is true, etc. If the conditions are implemented as nested if-statements
we will have tests that simultaneously have the first and the second condition as, say, true.  Thus, condition coverage requirements do not result in a partition.


Practicality, covering the criterion $\{C(i)\}_{i\in I}$, $C(i) \subseteq \Sigma^a$ may require excessive resources  that are not available to testers.  The following definition formalizes a common practice testers use to overcome this situation. Specifically, it formalizes the notion of relaxing coverage requirements:

\begin{definition}[coverage relaxation]
A coverage criterion $\{C^{'}(i)\}_{i\in J}$, $C^{'}(i) \subseteq \Sigma^a$ is a relaxation of a coverage criterion $\{C(i)\}_{i\in I}$, $C(i) \subseteq \Sigma^a $ if $J$ is a partition of $I$ and if for $j \in J$, $j = \{i_1, \ldots, i_k | i_l \in I, l = 1, \ldots, k\}$ then $C^{'}(j) = C(i_1) \cup \ldots \cup C(i_k)$.
\end{definition}

Clearly, $|J| \leq |I|$ is in the above definition, thus reducing the resources required to achieve the relaxed coverage criterion.
Consider code coverage, for example:
When full coverage of all lines of code is impossible, people usually settle for covering at least one line of code for each function. In our context, this translates to a partition of $I$ as defined above. Yet another relaxation example is found in synchronization coverage \cite{DBLP:conf/ppopp/BronFMNU05}. In synchronization coverage, there are coverage requirements $C(l, p^1_h, p^2_h)$ for each lock l, and each program location $p^1_h$ and $p^2_h$ that attempts to hold the lock l.  The coverage requirement $C(l, p^1_h, p^2_h)$ requires a sequence of events in which lock l is held by $p^1_h$ and then program location $p^2_h$ attempts to hold the lock.  Due to the availability of testing resources, the tester may want to relax this coverage requirement by only requiring some sequence of events in which any program location holds lock l. Then another program location attempts to obtain the lock l.  This will result in a new coverage requirement C(l) in which $C(l) = \bigcup_{p^1_h, p^2_h} C(l, p^1_h, p^2_h)$, which is a relaxation of the original synchronization coverage.  In that way,  coverage relaxation lets the testers control the testing effort.

\subsection{Coverage Criteria Examples}
\label{sec:approach:examples}
In this section, we motivate our definitions by showing that they naturally extend existing coverage notions and how they allow for formal specifications of best practices. We will later examine specific use cases in \autoref{sec:evaluation}.

We first show how our definitions generalize the classic $t$-way coverage on finite test spaces. Such coverage models have gained wide usage in the industry and are generally known as \textit{Combinatorial Test Design}~\cite{CTD}.

\begin{example}[classic $t$-way coverage] 
The classic $t$-way coverage is defined on finite spaces, i.e., when $P  \subseteq \Sigma^n$ for $n \in \mathbb{N}$. In this case, for a small $\,t \leq n$ and for each choice of $t$ indexes $i_1, \ldots, i_t, 1 \leq i_l \leq n$, and $t$ letters $\sigma_1, \ldots, \sigma_t \in \Sigma^t$, the coverage criterion is defined by
$C(i_1, \dots, i_t, \sigma_1, \dots, \sigma_t)=$ $\{ T \in \Sigma^n \colon T[i_k] = \sigma_k \text{ for } k=1,\ldots,t  \}$.
\end{example}

The following examples motivate the above definitions, focusing on test design in which the order of occurrences of events is essential. We outline how the approach generalizes Kuhn and Higdon's sequence coverage criterion~\cite{kuhn2010practical}, starting by demonstrating how to express Kuhn and Higdon's criterion using our framework. 

\begin{example}[Kuhn and Higdon's coverage criterion]\label{example:kuhn} Under \autoref{def:cov_crit}, Kuhn and Higdon's definition of $t$-sequence coverage~\cite{kuhn2010practical} is obtained by taking $I = \Sigma^t$ and $C({\sigma_1\cdots \sigma_t}) = \mathcal{L}(\Sigma^* \sigma_1 \Sigma^* \cdots \Sigma^* \sigma_t \Sigma^*)$, where $\mathcal{L}$ is the set of words (language) defined by the regular expression.
\end{example}

In addition to Kuhn and Higdon's coverage criterion, our framework allows for defining richer criteria.
Indeed, we can consider criteria that are not precisely $t$-sequence coverage. We can, for example, require that $\sigma_i$ does not appear unnecessarily, but only once. This leads to the following customized definition. 

\begin{example}[Like Kuhn and Higdon's coverage criterion] Take $I = \Sigma^t$ and $C({\sigma_1\cdots \sigma_t}) = \mathcal{L}(\Sigma'^* \sigma_1 \Sigma'^* \cdots \Sigma'^* \sigma_t \Sigma'^*)$ where $\Sigma' = \Sigma \setminus \{\sigma_1,\dots, \sigma_t\}$.
\end{example}

For example, with a SUT domain in mind, we would like to test all permutations of events in which a message is sent before receiving it. 

\begin{example}[message order]
Consider a language $\Sigma$ that contains subsets of send messages events $S \subseteq \Sigma$ and subsets of receive messages events $R \subseteq \Sigma$.  We then consider 
$C(s, r)=  \{ T \in \Sigma^n \colon T[i] = s, T[j] = r \text{ for  some }\allowbreak 1 \le i < j \le n \},\allowbreak s \in S, r \in R$
as our set of coverage requirements.  
Note, of course, that an event can be either ``send", ``receive," and maybe even ``other," but $R\cap S=\emptyset$. 
\end{example}

Such customized coverage criteria can be derived from known concurrent bug patterns or specific test concerns. Practice shows that this improves the quality of testing~\cite{1213511}.

A more concrete SUT will lead to further customization of coverage requirements. Consider for example a bank customer with two accounts. A transaction may reduce $m$~dollars from the first, and another may increase the second account by $m$~dollars. We are concerned that inconsistency may arise if the transaction is stopped in the middle. This leads to the following customized coverage requirement.  

\begin{example}[transaction safety]
We define $\Sigma$ as the set of possible transactions, $D \subseteq \Sigma$ as the set of transactions that reduce money, $A \subseteq \Sigma$ as the set of transactions that add money, and $\Sigma^{'} = \Sigma - (D \cup A)$. We capture our test concern by $C({d,a}) = \mathcal{L}(\Sigma^{'}d \Sigma^{'} a\Sigma^{'}), d \in D, a \in A$.
\end{example}

Many times coverage is motivated by symmetries. When we are required to see event $\sigma_1$ and then event $\sigma_2$, we implicitly claim that all permutations of events occurring before $\sigma_1$ are equivalent for the purpose of revealing a problem which is basically a symmetry claim (we also argue that all permutations between $\sigma_1$ and $\sigma_2$ as well as all permutations after $\sigma_2$ are equivalent).  Our framework supports the utilization of expected SUT symmetries in ways other than symmetries on execution order, as illustrated in the following example.        

Suppose we would like to check that a tic-tac-toe game-playing application works. There are 255,168~ways to play a game, i.e., put alternate Xs and Os on free squares in the $3 
\times 3$ board, where the game ends when a player wins or all cells are marked (tie). Nevertheless, if we assume the algorithm should work the same for symmetric boards (under eight rotations and reflections), we can considerably reduce the test space to~$26,830$.

\begin{example}[using symmetries to test an $n \times n$ board game]
\label{example:ttt}
Formally, the set, $P$, of equivalence classes of sequences of game-plays in an $n \times n$ game is $P=n^2!/{\sim}$ where $\sim$ is an equivalence relation over the eight rotations and reflections. In our setting, a coverage criterion is an equivalence class: $C([w])= [w]$. We use the common notations where $w$ is a game sequence and $[w]$ is the equivalence class that $w$ belongs to. 


To illustrate the construction of $C([w])$, we refocus on tic-tac-toe with $n=3$, which means $n^2=9$ squares.  We notice that the first move has only three possibilities (up to symmetry): center, corner, or center of an edge. If we number the places as $1,2,\ldots,9$ as shown in \autoref{TTT}, then the first move can only be from the symmetry equivalence classes [1], [2], or [5].  Concretely, a move from [1] could be 1, but it also could be 3, 7, or 9. As we assume that on each equivalence set, the algorithm either makes a mistake or not, then it is enough to test, say 7, to test a game with the first movement coming from $[1] = \{1, 3, 7, 9\}$.  We will relax this assumption in \autoref{sec:approach:bayesian} by introducing a Bayesian approach representing our confidence that an equivalence class does not contain a bug. 

\begin{figure}
    \centering
    \begin{tabular}{c|c|c}
         1 & 2 & 3 \\
         \hline
         4 & 5 & 6 \\
         \hline
         7 & 8 & 9 
    \end{tabular}
    \caption{A Tic-Tac-Toe board.}
    \label{TTT}
\end{figure}

To construct the equivalence classes $C([w])$, we apply the symmetries inductively to the game's evolution. The above symmetry application on the first move of the game reduces the space to be tested from $9!$ to $3\cdot 8!$, which is only $1/3$ of the space.
Furthermore, we can reduce it even further to $8\cdot 7!+16\cdot 6!$ (which is only $1/7$ of the space) by noticing that:
\begin{enumerate}[leftmargin=*]
    \item after playing 1, the next move can only be 2,3,5,6, or 9.  
    \begin{enumerate}[leftmargin=*]
        \item after playing 1,5 only 2,3,6, or 9 can be played.
        \item after playing 1,9 only 2,3,5, or 6 can be played.
    \end{enumerate}
    \item after playing 2, the next move can only be 1,4,5,7, or 8.
    \begin{enumerate}[leftmargin=*]
        \item after playing 2,5 only 2,3,6, or 9 can be played.
    \end{enumerate}
    \item after playing 5, the next move can only be 1 or 2.
    \begin{enumerate}[leftmargin=*]
        \item after playing 5,1 only 2,3,6, or 9 can be played.
    \end{enumerate}
\end{enumerate}

We proceed this way to define an equivalence set over the sequence of entire game plays, which define the equivalent classes $C([w])$. Note that the permutation requirement can be encoded by having a finite automaton with $9!$~states. Adding the constraints above is straightforward.

Take $I=n^2!/\sim$ and $C([w])= [w]$ where $\sim$ is an equivalence relation of the eight rotations and reflections, $n^2!/\sim$ denotes the set of equivalence classes of this relation and $[w]$ is the equivalence class of $w$.
\end{example}

\subsection{A Bayesian Risk-Reduction Approach}
\label{sec:approach:bayesian}
Exhaustive execution of $P$ is hard or even impossible as $P$ may be infinite.  If $I$ is finite, executing representatives from $C(i)_{i \in I}$ is a feasible alternative to exhaustive execution of $P$, though even that may be too hard due to the size of $I$ and difficulties in generating elements of $C(i)_{i \in I}$. Another challenge is that even if we execute a test in $C(i)$, we are not guaranteed, in general, that there are no bugs that can be exposed by executing another test in $C(i)$. There are several possible reasons for that:

\begin{enumerate}
\item The definition of $C(i)_{i \in I}$ is based on the tester's domain knowledge that may mistake. For example, the assumption in the tic-tac-toe example (\autoref{example:ttt}), namely, that mistakes are invariant under the eight possible symmetries, may be wrong. Thus, we may expose a problem by playing the upper left corner and then the center while not revealing it when playing the upper right corner and then the center.

\item In practice, we do not control all of the SUT's inputs. Specifically, it is difficult to control and specify the environment configuration and state. For example, a test that writes to a file may control whether or not the file exists and that we have written permission. However, the level of the operating system or the state of the memory garbage collector may not be a parameter controlled by the test. As a result, our confidence of not having a problem due to the execution of an element in $C(i)$ only increases when repeatedly running elements in $C(i)$.
\end{enumerate}

We thus apply a Bayesian approach that quantifies the risk of having a bug in the SUT given the tests that were executed so far. We also use a risk measure to guide the generation of additional tests to best decrease the risk of having a bug in the SUT. In the description below, we do not incorporate bug prediction information as was done in~\cite{9697382}.  If available, the method can be extended to incorporate bug prediction information by modifying the priors.

We define an indicator random variable $X_{C(i)}$.  For a given test $t$, $X_{C(i)}(t) = 1$ if $t \in C(i)$ and 0 otherwise. We first discuss the case in which $\{C(i)\}_{i \in I}$ is a partition. We randomly choose a test $t$, execute it, and determine if the test succeeded and to which $C(i)$ $t$ belongs. Given that $Y$ is the indicator variable of not having a bug, we model $P(Y|X_{C(i)}=1)$ as a Bernoulli distribution with a beta conjugate prior initialized to the uniform beta distribution (both $\alpha$ and $\beta$ parameters are set to one, where $\alpha$ is the weight representing the average success of the test $t$ when executed on $C(i)$).  Bayes rule is used to update the beta prior for each $C(i)$ as we gather evidence that $C(i)$ does not contain a bug. The update rule is simple; increase $\alpha$ by one each time the test succeeds and increase $\beta$ by one if it fails.
When the SUT is corrected based on the detected bugs, $\alpha$ and $\beta$ are reset to 1 for each $C(i)$, and the process repeats. If $\{C(i)\}_{i \in I}$ is not a partition an evidence of a test $t$ execution result may apply to more than one $C(i)$ and we can update the conditional probabilities estimation for each one of them.

To produce an overall estimation of the likelihood of a bug, $P(X_{C(i)} = 1)$ can be estimated using some profile of the software usage (either collected empirically or estimated). After running $k$ chosen tests,
$\sum_{i \in I} P(Y|X_{C_{i}} = 1)P(X_{C(i)} = 1)$ is used to update the likelihood of a bug. Here we use the assumption that $C(i)$ is a partition and we use the current beta prior associated with each $C(i)$ to estimate  $P(Y|X_{C_{i}} = 1)$.  We can always obtain a partition by refining the set of coverage criteria, $\{C(i)\}_{i \in I}$, using intersections.  In addition, if $\{C(i)\}_{i \in I}$ is not a partition, the likelihood function above is still relevant but no longer has a probability interpretation.  Instead, it serves to measure more than once evidence that applies to intersections of $C(i)$s.  If bugs in each $C(i)$ are associated with a different loss $l(C(i))$, we can represent the average expected loss of the system as $\sum_{i \in I} l(C(i)))P(Y|X_{C_{i}} = 1)\allowbreak P(X_{C(i)} = 1)$.

Another way of handling the case in which $C(i)_{i \in I}$ is not a partition is to focus on the same random variables as above and apply Bayesian-network discovery techniques~\cite{10.5555/1005332.1005352} to learn the edges in the network. The network is learned from a profile of the system usage.  More specifically, we consider the joint distribution of the variables $X_{C_{i}}, i \in I$ and obtain N samples of the joint distribution from the SUT profile.  Assuming here that $I$ is small, one may use the Bayesian scoring criterion described in chapter eight of \cite{Neapolitan2004Learning} to choose the Bayesian network that best fits the $N$ samples.  

The above risk and loss functions can guide the test generation process. In general, we prefer to take steps in the test generation process that decrease the risk of finding the bug or the expected average loss from finding it. The process of generating a test $t$ reaching $C(i)$ may include the realization of several conditions. For example, we need to read and then write to some shared resource to cover $C(i)$. Facing several generation alternatives of the tests in $P$, we may prefer to realize conditions that are needed in order to reach $C(i)$ over conditions that are needed to reach $C(j)$ if $P(Y|X_{C(i)} = 1)  > P(Y|X_{C(j)} = 1)$.  This includes the special case in which $C(i)$ was never visited, which will be preferred over $C(j)$s that were visited many times. In addition, if the loss function is given and the loss associated with a defect in $C(i)$ is high, we may prefer revisiting $C(i)$, even if the probability of finding a bug in $C(i)$ is low. 

Note that the case in which no bugs are found in the above procedure only indicates that the current abstraction level imposed by the set of coverage criteria $C(i), i \in I$ is no longer effective in finding bugs in the SUT. It does not mean the system has no bugs. Indeed, as a tester, if my expert opinion is that the system still has defects, I should probably attempt to design, following our methodology, additional coverage criteria that need to be covered. On the other hand, if my expert opinion as a tester is that the set of coverage criteria $C(i), i \in I$ is sufficient, then it is consistent on my part to stop the testing if the above search procedure no longer yields bugs. 

\section{A Tool-Driven Methodology}
\label{sec:tool}
This section demonstrates how our approach can be applied in practice, demonstrating the coverage criteria specification and the test suite generation. To support this process and test the performance of generalized coverage criteria in the real world, we added a tool that allows users to generate small test suites (sets of tests) from their programs. This tool uses a ranking function provided by the user and applies a genetic algorithm (GA)~\cite{eiben2003introduction} to construct test suites with high ranks (i.e., individuals represent test suites). The full implementation details are presented in \autoref{sec:eval:generation}.
The tool and code examples are at \url{github.com/bThink-BGU/Papers-2023-TSE-Sequence-Testing}. 

\subsection{Mapping from 
Theoretical to Implementation Terms}
The proposed tools are designed to support the methodology described in \autoref{sec:approach}. While \autoref{def:cov_crit} only specifies a Boolean condition of coverage (i.e., a criterion is either covered or not), our tool, for practical reasons, allows for a quantitative measure of coverage. 

Specifically, we propose counting the number of $C(i)$s that a test suite covers and maximizing it as much as possible. We note that our tool only gives a sub-optimal solution. The optimization process can be controlled by tweaking the parameters to get the required balance between computation resources and quality. 

The terms map as follows:
\begin{itemize}
\item The test model $P$ is a set of 50k possible executions of the system-behavior model (i.e., paths).

\item The test suite $S$ is generated by the tool. Its size can be changed according to the available resources. 

\item The coverage criteria $C=\{C(i)\}_{i\in I}, C(i) \subseteq \Sigma^a$
is specified as a ranking function that takes a test suite $S$ and returns the number of $C(i)$s that are covered. 

\item The coverage ratio $\Gamma_C(S, P)$ (\autoref{def:cov_rank}) is computed by dividing the result of the ranking function by all possible sequences in $C$. For example, if there are 11 possible events, a 2-way ranking function of $P$ is bounded in $11^2=121$.
\end{itemize}

\subsection{Coverage-Criteria Specification}
\label{sec:tool:coverage}
Our test-suite-generation tool uses a user-defined ranking function to determine the quality of the test suites. This function takes a test suite, represented as a set of lists of events, and returns a number that models how well the suite covers the criteria that the user is interested in. In \autoref{sec:evaluation}, we measure different coverage criteria using this tool. We used, for example, this tool to express our interest in counting how many different regular expressions of the form $\Sigma^* \sigma_1 \Sigma^* \sigma_2 \Sigma^*$ are matched by at least one test in the suite, where $(\sigma_1,\sigma_2)$ is a pair of events. This is, of course, Kuhn and Higdon's 2-way sequence coverage criterion, as presented in \autoref{lst:ranking_code}. The ranking function is an implementation of coverage criteria. Searching for top test suites according to the coverage criteria means that the chosen test suite maximizes the coverage of the test-case space. The ranking function attaches value to each candidate test suite according to the suite's opportunity for system coverage. 

\begin{lstlisting}[
  float=tpb,
  language=java,
  label={lst:ranking_code},
  caption={A ranking function in Java that implements Kuhn and Higdon's 2-way coverage criterion. },
]
private long twoWayRank(Set<List<String>> suite) {
  var allPairs = new ArrayList<>();
  for (var test : suite)
    for (var i = 0; i < test.size() - 1; i++)
      for (var j = i + 1; j < test.size(); j++)
        allPairs.add("(" + test.get(i) + "," + 
                      test.get(j) + ")");
  return allPairs.stream().distinct().count();
}
\end{lstlisting}

\subsection{Test-Suite Generation}
\label{sec:tool:generation}
To extract test suites with a high rank from the BP model, we developed a tool based on a genetic algorithm (GA)~\cite{eiben2003introduction} that evolves test suites. The tool enables the testing engineer the freedom to tailor the parameters required to build good test suites for the system. The tester may adopt the following characteristics: the size of the pool of valid test cases, the test-suite size, the ranking function, and the search algorithm. 
Given these parameters, the tool creates a vast pool of valid test cases and finds test suites with a high ranking. 

Specifically, for the GA and the genetic operators, we use Jenetics---a Java library for evolutionary algorithms~\cite{wilhelmstotter2022jenetics}. Our tool creates a pool of valid test cases by performing 50K random walks on the given system-behavior model. It then utilizes GA to search for highly ranked test suites. Each individual, i.e., a test suite, is represented as a set of $n$ tests, where different $n$ values can be used. The initial population is randomly created from the pool of valid test cases. The fitness function is the ranking function provided by the user. We apply a standard mutation operator that replaces each test in each individual with another random test, with a probability of 0.05. We used a partially matched crossover~\cite{eiben2003introduction} with a distribution of 0.7 to avoid repetitions of tests within a test suite. The complete evolutionary hyperparameters are summarized in \autoref{tab:setup}. Notably, while the hyperparameters may be further optimized, depending on the model and coverage criteria, they were selected as they proved robust to many models and coverage criteria.

In \autoref{sec:evaluation}, we evaluate the sensitivity of the various parameters to the quality of the obtained results.

\begin{table}
\caption{Evolutionary hyperparameters.}
\label{tab:setup}
\centering
\begin{tabularx}{\columnwidth}{rX}
\toprule
Representation & \lstin{Set<List<String>>} \\
Mutation &  Uniform test replacement \\
Recombination & Partially Matched Crossover   \\
Mutation probability & 0.05\\
Recombination probability & 0.7\\
Parent selection & Tournament with $k=3$\\
Survivor selection & Generational replacement\\
Population size & 100\\
Termination criteria & 300 generations or best-fitness\newline convergence ($\epsilon \leq 0.001$)\\
\bottomrule
\end{tabularx}
\end{table}

\section{Evaluation}
\label{sec:evaluation}
\autoref{sec:tool} provided a qualitative assessment of our modeling approach, demonstrating its applicability for testing. For example, we showed how coverage criteria could be specified as ranking functions in JavaScript.

Here, we have focused on a quantitative evaluation of our approach. Specifically, we evaluate here: 1) the effectiveness of different coverage criteria in detecting bugs; 2) the efficiency of our GA-based tool, and 3) the validity of our Bayesian risk-reduction approach. To this end, we begin with the \textit{alternating-bit protocol}, which is a standard benchmark for formal verification and modeling~\cite{chukharev2020sat}. Next, we evaluate our approach on a web application called \textit{Moodle}, which is a popular, open-source learning management system. The use case of Moodle demonstrates the applicability of the method to real-life testing.

For specifying the system-behavior model, we used the \textit{Behavioral Programming} modeling paradigm~\cite{Harel2012BehavioralProgramming2}, designed for specifying behavior in a natural and intuitive manner that is aligned with how users perceive the requirements of a system. We begin with a short description of this paradigm and a demonstration of its usefulness for specifying system-behavior models.

The systems under test, the testing model, and execution instructions are in our repository at \url{github.com/bThink-BGU/Papers-2023-TSE-Sequence-Testing}.

\subsection{Behavioral Programming}
In behavioral programming (BP), a user specifies a set of scenarios that may, must, or must not happen. Each scenario is a simple sequential thread of execution and is thus called a \emph{b-thread}. B-threads are typically aligned with system requirements, such as ``user must log in before using the system,''  or ``every file-read action must be preceded by a file-open action,'' etc. The set of b-threads, called behavioral program (\emph{b-program}), specifies the overall system behavior and, in our case, what needs to be tested. At run-time, all b-threads participating in a b-program are combined, yielding a complex behavior that is consistent with all the b-threads. 

To synchronize the b-threads behaviors, Harel et al.~\cite{Harel2012BehavioralProgramming2} proposed a simple b-thread integration protocol. The protocol consists of each b-thread submitting a statement before selecting an action to perform, where actions are represented as events. The statement declares which events the b-thread requests, which events it waits for (but does not request), and which events it blocks (forbids from happening). After submitting the statement, the b-thread is paused. When all b-threads have submitted their statements, we say the b-program has reached a \emph{synchronization point}. Then, a central event arbiter selects a single event that was requested and was not blocked. Having selected an event, the arbiter resumes b-threads that requested or waited for that event. The rest of the b-threads remain paused, and their current statements are used in the next synchronization point. 

From a formal point of view, BP semantics are typically defined in terms of transition systems, where each b-thread is a labeled transition system (LTS), and the execution engine generates a cohesive LTS on the fly~\cite{Elyasaf2020COBP, Harel2010ProgrammingCoordinated}. In this paper, we use an implementation of BP, called BPjs~\cite{BarSinai2018BPjs}, where b-threads are specified using simple JavaScript code snippets (hence the name), and the integration mechanism is developed in Java using the Rhino library~\cite{rhino}. We use BPjs to specify the test model, generate the cohesive LTS that models all possible sequences of events that test the system and execute the test model (which is equivalent to performing a random walk on the cohesive LTS). As described next, since all of these sequences are usually huge, we need to sample them wisely.

BP has been used for different development phases, such as requirement specification and solicitation~\cite{Elyasaf2018LSC4IoT, Elyasaf2019RoboCup2, Poliansky2022GPBP}, implementation~\cite{Harel2010ProgrammingCoordinated}, and verification~\cite{Harel2011ModelcheckingBehavioral2}, and also for testing~\cite{BarSinai2019Satellite, Greenyer2021ScenarioBasedMA, wiecher2021integrated, weiss2021testing}. 

To understand the concepts of behavioral programming and how it can be used for specifying system-behavior model, we present a benchmark example of Bombarda and Gargantini~\cite{bombarda2020automata}, showing how the regular expression they proposed can be modeled using BP. 
This is an example of a vault that can be unlocked by the combination ``12345''. The code in \autoref{lst:vault} specifies two b-threads, one for pressing the keys and one for checking the combination. While this small example can be modeled using a single b-thread, breaking the specification into two modules allows us to align each b-thread to a different testing requirement. The first b-thread is aligned with the requirement that the safe has a keypad with nine digits that can be pressed in any order. It continuously requests pressing any of these keys. The second b-thread is aligned with the requirement that the safe is opened only when the correct code is dialed. While it does not request keys, it waits for the correct sequence and then requests the \lstin{Open} event while blocking all other events (i.e., keys). At run-time, the b-threads are executed simultaneously and synchronized using the protocol described before.

\begin{lstlisting}[
  float=tb,
  label={lst:vault},
  caption={The vault example~\cite{bombarda2020automata} implemented with BP. The program has two b-threads --- the first continuously requests to press any of these keys. The second b-thread waits for the correct key sequence and then opens the safe while blocking any other action.},
]
const code = '12345'
const AnyKey = [ Event('1'), $\dots$, Event('9') ]

bthread('press keys', function() {
  while(true)
    sync({ request: AnyKey }) //press a random key
})

bthread('check code', function() {
  for(let i=0; i<5; i++) {
    let key = sync({ waitFor: AnyKey })
    if (code[i] != key) //wrong vault code
      sync({ block: bp.all }) 
  }
  //correct vault code
  sync({ request: Event('Open'), 
           block: Event('Open').negate() })
  sync({ block: bp.all }) 
})
\end{lstlisting}

\subsection{Alternating-Bit Protocol}
The alternating-bit protocol (ABP)~\cite{bartlett1969note} is a full-duplex communication protocol using an unreliable communication channel. Each packet in this protocol is repeatedly sent until the sender receives an acknowledgment from the client. Each packet is attached with a single bit of metadata that indicates the correct order of the messages. This bit alternates in each packet. 
When a packet is received, the receiver verifies that the attached bit is correct and sends back an acknowledgment message with the same bit attached. 

In order to assess our methodology, we implemented the ABP protocol as described in~\cite{fMBT}. Our implementation involved translating the model to Java for convenience and utilizing it as the system under test. Additionally, we introduced a variety of defects to investigate different facets of our approach, which are explained in detail below.

In addition to the SUT implementation, we created a testing model using BP that also includes ``rainy-day'' scenarios. Specifically, the model specifies two types of noises that can occur in both communication directions (sender to receiver and vice versa). The noises are a loss of a message and a change in the order of the messages. Both the SUT and the testing model are in our repository.

We applied a ``white-box'' testing approach where the testing model is used to drive the SUT and check that it follows the protocol. Each test case (list of events generated from the b-program) triggers the SUT to act and execute the event action. For example, the \lstin{send} event triggers the SUT to send a data packet. An instrumentation layer checks if the conditions to execute the event are met, and then the action is carried out, and the internal state of the SUT updates accordingly. If the conditions for triggering the event are not met by the SUT, an error occurs, and the test fails. The test succeeds if all events execute successfully. We represented some actions using two events to allow a high granularity in the test cases. For example, when a receiver sends an acknowledgment message, it may be either correct (event \lstin{rAck}) or incorrect (\lstin{rNak}). Similarly, the sender may receive this message correctly (\lstin{sAck}) or incorrectly (\lstin{sNak}).

\subsubsection{Coverage-Criteria Evaluation}
\label{sec:eval:criteria}

\autoref{sec:approach:examples} showed how testers could specify coverage criteria that generalize the notion of coverage and help in focusing the test efforts. Here we evaluate the effectiveness of coverage criteria in detecting different bugs.
To this end, we injected four bugs in the SUT as follows:

\begin{itemize}
    \item{sAck,sAck:} When the sender receives two acknowledgments in a row, it disregards them both and re-sends the message. 
    \item{rNak,rAck:} When the receiver has to send an acknowledgment after a not-acknowledges, it does not send an acknowledgment.
    \item{sNak,sNak,rAck:} When the receiver has to send an acknowledgment after the sender receives two following not-acknowledges, it does not send an acknowledgment.
     \item{send,send,sAck:} When the sender receives an acknowledgment after sending the following message twice, it disregards the acknowledgment and re-sends the last message.
\end{itemize}

Next, we used the BP-based test model to generate many test cases, collect them into test suites, and examine how often they find the injected bugs in the SUT. We ran the model 50K times and generated 50K valid test cases. We built test suites out of these test cases, each has ten test cases. Our challenge was finding the best test suites with the highest probability of catching the injected bugs. To this end, we defined two ranking functions and used our GA-based tool to find test suites with a high rank.

The first ranking function is based on Kuhn and Higdon's method that counts all sequences of $n$ events but not necessarily consecutive $n$ events (i.e., $\{\Sigma^* \sigma_1 \Sigma^* \sigma_2  \Sigma^* \colon \sigma_1, \sigma_2 \in \Sigma \}$ or its equivalent for three letters bugs). The second-ranking function counts all $n$ consecutive events in a test suite (i.e., $\Sigma^* w \Sigma^*$). Our thesis is that maximizing the consecutive events maximizes the probability of catching the bug, i.e., that the second-ranking function is better for our purposes. The point that we are trying to make here is that there are situations where our generalization that allows ranking functions other than Kuhn and Higdon's criterion has real usage. For a baseline, we also used a `random' method for the  test-suite generation that simply peeks ten test cases at random. 

For each method and an injected bug, we counted how many tests detected the bug. We repeated this process 1K times and averaged the results. The results are summarized in \autoref{tab:probabilities}.
The second column specifies the sequences of events that trigger the injected bugs. 
For example, the sequence sNak, sNak, rAck represents the case where the sender receives twice an acknowledgment message with the wrong bit, and then the receiver sends the correct acknowledgment message.
As the random column suggests, some bugs are frequent, and some occur in rare corner cases.

The results show that the generalized criterion $\Sigma^* w \Sigma^*$ is always better and that it is much better than Kuhn and Higdon's criterion when the bug is rare. 
This criterion model a heuristic that applies to many systems where specific sequences of consecutive events trigger bugs. We call it a `generalized' criterion because it demonstrates how developers can generalize Kuhn and Higdon's approach for modeling new heuristics that target new types of bugs. Note that we are not claiming that our criterion is better than Kuhn and Higdon's criterion in general, only that generalizing allows us to target specific types of bugs better.

\begin{table}
\caption{The probability of catching bugs in the alternating-bit protocol and Moodle LMS, for each test-suite generation method. The unranked method randomly generates test suites. $w$ represents a sequence that triggers the bug.}
\label{tab:probabilities}
\centering
\begin{tabular}{@{\hskip 3px}l@{\hskip 6px}p{85px}@{\hskip 4px}|P{33px}P{25px}@{\hskip 6px}c@{\hskip 3px}}
\toprule
Domain & $w$ & Unranked\newline (Random) & Kuhn-\newline Higdon & $\Sigma^*w\Sigma^*$ \\ \midrule\midrule
\multirow{4}{*}{Alt. Bit} & sAck, sAck &   0.219     &   0.198   & 0.709 \\
& rNak, rAck  &  0.732      & 0.643 & 0.988 \\
& sNak, sNak, rAck &     0.067 &  0.091 &  0.216 \\
& send, send, sAck &  0.821 & 0.806     & 0.978 \\ \midrule
Moodle & teacher.AddQuestion, student.SubmitExam, teacher.SubmitQuestion    &  0.372 & 0.986 & 0.512 \\
\bottomrule
\end{tabular}
\end{table}

\subsubsection{Test-suite Generation Evaluation}
\label{sec:eval:generation}
To evaluate our GA-based tool for generating test suites, we compare it to two other methods for different coverage criteria and different suite sizes. To produce the test suites, we first created a pool of 50K valid test cases extracted from our model. We then used three methods for generating test suites from the pool ($n$ is the test suite size): 
\begin{enumerate}
\item \textit{Unranked:} Randomly peeking $n$ test cases.

\item \textit{Best of 1K:} Repeating the `unranked' process 1K times and choosing the suite with the highest rank.

\item \textit{GA:} Our GA-based tool described in \autoref{sec:tool:generation}. 
\end{enumerate}

We check each method with the same coverage criteria as before (Kuhn and Higdon and $\Sigma^* w \Sigma^*$) and three test-suite sizes: 5, 10, and 20. For each test, we measure the ranking function value (called `rank') and the run-time (`time') in seconds. The results presented in \autoref{tab:ranking}, show that GA works best in terms of time and ranking. Kuhn and Higdon's 2-way rank is constant since there are eleven possible events, and therefore the rank is bounded in $11^2=121$.

\autoref{fig:effGraph} depicts the average generation time for each method (i.e., the time row of the $\Sigma^* w \Sigma^*$ criterion in \autoref{tab:ranking}). 
This visualization clearly shows that GA outperforms other methods both in time and obtained rank in all three test sizes. The advantage grows with the size of the test.

\begin{table*}
\caption{The performance of test-suite generation methods in terms of quality and time, evaluated on the alternating-bit protocol and Moodle LMS. We examined the performance for each method (GA, ``best of 1K'', and unranked), possible test-suite size (5, 10, and 20), and scoring functions (Kuhn and Higdon and $\Sigma^* w \Sigma^*$). Numbers are averaged over 1K runs.}
\label{tab:ranking}
\centering
\begin{tabular}{lp{25px}l||ccc|ccc|ccc}
\toprule
\multicolumn{3}{r||}{Method} & \multicolumn{3}{c|}{GA} & \multicolumn{3}{c|}{Best of 1K} & \multicolumn{3}{c}{Unranked (Random)} \\
\multicolumn{3}{r||}{Suite size} & 5 & 10 & 20 & 5 & 10 & 20 & 5 & 10 & 20 \\ \midrule\midrule

\multirow{6}{*}{Alt. Bit} & \multirow{3}{25px}{Kuhn-\newline Higdon} & Time (mSec) & $78.1$ &$88.7$  & $125$ & $622.4$ &$859.4$  & $810$ &$0$   &$0$  & $0$ \\
& & Rank (max) & $121$ &$121$  &$121$  & $121$ &$121$  &$121$  & $121$ &$121$  &$121$  \\
& & Rank (mean) & $120.457$ & $120.797$ & $120.879$ & $120.879$ & $120.879$ & $120.879$ & $112.983$ & $117.496$ & $118.919$ \\ \cmidrule{2-12}

& \multirow{3}{*}{$\Sigma^*w\Sigma^*$} & Time (mSec) & $31.2$ &$38.9$  & $38.9$ & $519.3$ &$528.9$  & $561$ &$0$   &$0$  & $11$ \\
& & Rank (max) & $41$ &$58$  &$78$  & $39$  &$55$  &$62$  & $34$ &$42$  &$50$  \\ 
& & Rank (mean) & $37.579$ & $54.162$ & $63.801$ & $36.249$ & $45.915$ & $53.572$ & $25.575$ & $32.61$ & $40.325$ \\ \midrule\midrule

\multirow{6}{*}{Moodle} & \multirow{3}{25px}{Kuhn-\newline Higdon} & Time (Sec) & $177.433$ &$299.626$  & $474.957$ & $476.969$ &$945.588$  & $1757.526$ &$0.863$   &$1.987$  & $3.68$ \\
& & Rank (max) & $700$ &$700$  &$700$  & $696$ &$700$  &$700$  & $673$ &$696$  &$699$  \\
& & Rank (mean) & $684$ & $695$ & $699$ & $673$ & $692$ & $699$ & $611$ & $647$ & $668$ \\ \cmidrule{2-12}

& \multirow{3}{*}{$\Sigma^*w\Sigma^*$} & Time (Sec) & $14.671$ &$15.353$  & $22.282$ & $21.158$ &$31.242$  & $49.068$ &$0.151$   &$0.135$  & $0.113$ \\
& & Rank (max) & $76$ &$111$  &$150$  & $71$  &$104$  &$139$  & $68$ &$97$  &$133$  \\ 
& & Rank (mean) & $68$ & $103$ & $141$ & $66$ & $98$ & $132$ & $57$ & $85$ & $118$ \\ \bottomrule
\end{tabular}
\end{table*}

\pgfplotsset{compat=1.3,width=0.8\columnwidth}
\begin{figure}
\centering
\vspace{3ex}
\begin{tikzpicture}
\begin{axis}[enlargelimits=0.1,
    xlabel={Time (mSec.)},
    ylabel={Rank},
    legend cell align=left,
    legend pos=north east,
]
\footnotesize
\addplot[
    scatter/classes={un={blue,mark=diamond*}, bf={red,mark=square*}, ga={purple,mark=triangle*}},
    scatter, mark=*, 
    scatter src=explicit symbolic,
    visualization depends on={value \thisrow{no} \as \Label} 
] table [meta=alg] {
time rank alg no
0   36  un 5
31   41  ga 5
519   40  bf 5
}
node[pos=0.0, xshift=-5px]{5} 
node[pos=0.08, yshift=6px]{5} 
node[pos=1.0, yshift=6px]{5}
;
\addplot[
    scatter/classes={un={blue,mark=diamond*}, bf={red,mark=square*}, ga={purple,mark=triangle*}},
    scatter, mark=*, 
    scatter src=explicit symbolic,
    visualization depends on={value \thisrow{no} \as \Label} 
] table [meta=alg] {
time rank alg no
0   42  un 10
39   58  ga 10
529   55  bf 10
}
node[pos=0.0, xshift=-7px]{10} 
node[pos=0.09, yshift=6px]{10} 
node[pos=1.0, yshift=6px]{10} 
;
\addplot[
    scatter/classes={un={blue,mark=diamond*}, bf={red,mark=square*}, ga={purple,mark=triangle*}},
    scatter, mark=*, 
    scatter src=explicit symbolic,
    visualization depends on={value \thisrow{no} \as \Label} 
] table [meta=alg] {
time rank alg no
11   50  un  20
39   78  ga 20
561   62  bf 20
}
node[pos=0.0, xshift=-7px]{20} 
node[pos=0.08, yshift=6px]{20} 
node[pos=1.0, yshift=6px]{20}
;
\legend{Unranked, Best of 1K, Genetic Algorithm}
\end{axis}
\end{tikzpicture}
\caption{Efficiency graph for the $\Sigma^* w \Sigma^*$ criterion, evaluated on the alternating-bit protocol. The graph displays each method's ratio between calculation and time to quality. The number of data points represents the test-suite size.}
\label{fig:effGraph}
\end{figure}
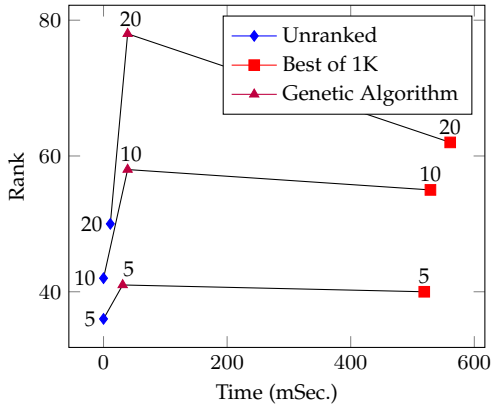

\subsubsection{Bayesian Risk-Reduction Evaluation}
\label{sec:eval:bayesian}
This section demonstrates how our Bayesian approach computes a stable estimation of the risk involved with each coverage criteria $C(i)$ when the system is continually tested and not fixed.

In our setting, there are $121$ coverage criteria, namely $11^2$ and we have run $t_1, \ldots, t_n$ tests, $n = 50{,}000$. When a test, $t_i$ is executed, it either hits a coverage criterion $C(i)$ or does not. Assuming that $C(i)$ is hit when $t_i$ is executed, we either find a defect or not, and the test either passes or not. Our focus is on the statistical dependency under a Bayesian setting of these two random variables. Discovering or failing to discover a defect during a test that targets some $C(i)$ triggers an update of the prior probability of finding a defect associated with $C(i)$. 
As the prior is a Beta function, it is controlled by two parameters $\alpha_i, \beta_i$ in which $E_i=\frac{\alpha_i}{\alpha_i+\beta_i}$ represents the estimated average of finding a defect if $C(i)$ is hit. The counters $\alpha_i$ and $\beta_i$ are continually updated using the Beta Bernoulli conjugate prior rule. Namely, if $C(i)$ was covered during the execution of $t_i$, we increment $\alpha_i$ by one if a bug was found and $\beta_i$ otherwise. We want to establish that the prior stabilizes over time, i.e., that $E_i$ converges when $n \to \infty$ and that the probability mass of the prior concentrates. The second is established if the variance of the Beta distribution, namely, $V_i =\frac{\alpha \beta}{(\alpha+\beta)^2(\alpha+\beta+1)}$ vanishes. We have run $50k$ tests and updated the priors for each coverage criterion, as explained above. 
We then removed all the $C(i)$s not covered by at least $1k$ tests and computed the maximal variance of the remaining $C(i)$s over time. Note that the current testing efforts are insufficient for the removed $C(i)$s, requiring additional test design and implementation. 
As depicted in \autoref{fig:max_variance}, the variance converges to zero as expected, indicating that our risk estimation gets very accurate after a reasonable number of tests.

\pgfplotsset{compat=1.3,width=0.7\columnwidth}
\begin{figure}
\centering
\begin{tikzpicture}
\begin{axis}
[grid = major, 
ylabel={Variance of risk estimator}, 
xlabel={Number of tests}, 
]
\footnotesize
\addplot[color=blue!50!red,smooth,tension=0.7,very thick,
] table [x index=0,y index=1,col sep=comma, ] {max_variance.csv}; 
\label{time}
\end{axis}
\end{tikzpicture}
\caption{The stabilization of our Bayesian-based risk estimation, evaluated on the alternating-bit protocol. The estimation of the risk of each coverage criteria $C(i)$ is stabilized when the system is continually tested and is not fixed. The graph shows that the variance of these parameters gets very small very quickly, indicating that our risk estimation gets very accurate after a reasonable number of tests.}
\label{fig:max_variance}
\end{figure}
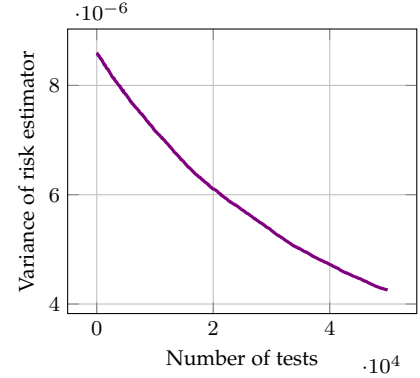

\subsection{Moodle LMS}
\label{sec:eval:moodle}
BP can be used for modeling systems that are not usually perceived as event-based. We demonstrate it on Moodle LMS --- a popular, open-source learning management system used by educators to create private websites with online courses to achieve learning goals~\cite{moodle}. While Moodle is developed as an object-oriented system, use it as a black box and model the interaction between the users and the system as an event-based system. 

In the supplemented material, we provide the specification of a BP-based model with three b-threads, each aligned to a different aspect of the system behavior, handled by a different type of user. The first b-thread specifies an administrator behavior that creates a course and enrolls users in it. The second b-thread specifies how an enrolled teacher adds a quiz with two questions to the course. Finally, the last b-thread specifies how an enrolled student waits for a question to be added and then answers the question. We used the Selenium WebDriver~\cite{selenium} for performing actions on Moodle UI. Using this full stack of b-program, actuators, Selenium WebDriver, and Moodle v3.9, we were able to run a few tests and detect the following bug.

According to Moodle documentation~\cite{moodle:quiz:faq}, a teacher cannot add questions to a quiz once a student attempts it. This behavior is enforced by the user interface (UI), as the ``Add question'' button disappears once a student attempts the quiz. Yet, if a student attempts the quiz while the teacher adds another question, the UI misbehaves and displays an exception. These are the exact steps that reveal the bug: (1) a teacher starts adding a second question to a quiz; (2) a student attempts to answer the first question; (3) the teacher submits the second question. 

\begin{table}[tb]
    \caption{Statistics on Moodle's b-program state space}
    \label{tab:moodle}
    \centering
    \begin{tabular}{c|c|c|c|c}
        \toprule
        B-threads & Events & States & Edges & Traces (possible tests) \\
        \midrule\midrule
        3 & 22 & 103 & 195 & 33,489 \\
        \bottomrule
    \end{tabular}
\end{table}

The Moodle example is far more complicated than the alternating-bit protocol. For the purpose of evaluating its complexity, we generated an automaton that represents the entire state space of our model. \autoref{tab:moodle} shows that although our model has only three intuitive b-threads and $22$ distinct events, the state space consists of $103$ states, $195$ edges, and $33{,}489$ distinct traces. Notably, the generated model includes only ``high-level'' events, such as \lstin{login(username, password)} or \lstin{enrollUser(course-id, username)}. To actuate the website using Selenium, we added ``low-level'' events to the model, such as \lstin{writeText(xpath, text)} and \lstin{click(xpath)} that receive an element unique path on the UI and perform actions on it. This extended model is too large to compute, though it does not raise a problem since it is executed directly without requiring the generation of the state space.

\subsubsection{Coverage-Criteria Evaluation}
\label{sec:eval:moodle:criteria}
Here we present how we evaluated the effectiveness of coverage criteria in detecting different bugs. While evaluating our approach on Moodle, we detected the aforementioned bug using Kuhn and Higdon's ``3-way'' criterion, emphasizing the usefulness of this criterion and its advantage. 
Since the SUT is a real system with real bugs, we did not have to inject bugs. Other than that, the evaluation method is the same as presented in \autoref{sec:eval:criteria}. 

The results are summarized in \autoref{tab:probabilities}.
The second column specifies the sequences of events that trigger the injected bugs. 
Specifically, the sequence \lstin{teacher.AddQuestion}, \lstin{student.SubmitExam}, \lstin{teacher.SubmitQuestion} represents the detected bug, as explained above.
According to the findings, Kuhn and Higdon's 3-way criterion is more effective in this case, which makes sense considering that the bug in question is dependent on three independent events, and the occurrence of other events has no impact on it. This indicates the necessity of a universal approach that does not replace current state-of-the-art methods but rather complements them with more generalized versions.

\subsubsection{Test-suite Generation Evaluation}
\label{sec:eval:moodle:generation}
Similar to \autoref{sec:eval:generation}, we compared our GA-based tool with two baselines, unranked and ``best of 1K'', measuring the value of the ranking function (called `rank') and the run-time (`time') in seconds over three test-suite sizes: 5, 10, and 20. 
The results are presented in \autoref{tab:moodle}, showing that GA works best in terms of time and ranking.

\autoref{fig:moodle:effGraph} depicts the average generation time for each method (i.e., the time row of the $\Sigma^* w \Sigma^*$ criterion in \autoref{tab:moodle}). 
Like in the alternating-bit protocol, GA is better than ``best of 1K'' both in time and obtained rank in all three test sizes. The advantage grows with the size of the test.

\pgfplotsset{compat=1.3,width=0.8\columnwidth}
\begin{figure}
\centering
\vspace{3ex}
\begin{tikzpicture}
\begin{axis}[enlargelimits=0.1,
    xlabel={Time (Sec.)},
    ylabel={Rank},
    legend cell align=left,
    legend pos=south east,
]
\footnotesize
\addplot[
    scatter/classes={un={blue,mark=diamond*}, bf={red,mark=square*}, ga={purple,mark=triangle*}},
    scatter, mark=*, 
    scatter src=explicit symbolic,
    nodes near coords*={\Label},font=\footnotesize,
    visualization depends on={value \thisrow{no} \as \Label} 
] table [meta=alg] {
time rank alg no
0   68  un 5
15   76  ga 5
21   71  bf 5
};
\addplot[
    scatter/classes={un={blue,mark=diamond*}, bf={red,mark=square*}, ga={purple,mark=triangle*}},
    scatter, mark=*, 
    scatter src=explicit symbolic,
    nodes near coords*={\Label},font=\footnotesize,
    visualization depends on={value \thisrow{no} \as \Label} 
] table [meta=alg] {
time rank alg no
0   97  un 10
15   111  ga 10
31   104  bf 10
};
\addplot[
    scatter/classes={un={blue,mark=diamond*}, bf={red,mark=square*}, ga={purple,mark=triangle*}},
    scatter, mark=*, 
    scatter src=explicit symbolic,
    nodes near coords*={\Label},font=\footnotesize,
    visualization depends on={value \thisrow{no} \as \Label} 
] table [meta=alg] {
time rank alg no
0   133  un  20
22   150  ga 20
49   139  bf 20
};
\legend{Unranked, Best of 1K, Genetic Alg.}
\end{axis}
\end{tikzpicture}
\caption{Efficiency graph for the $\Sigma^* w \Sigma^*$ criterion, evaluated on Moodle LMS. The graph displays the ratio between calculation and time to quality for each method. The number of data points represents the test-suite size.}
\label{fig:moodle:effGraph}
\end{figure}
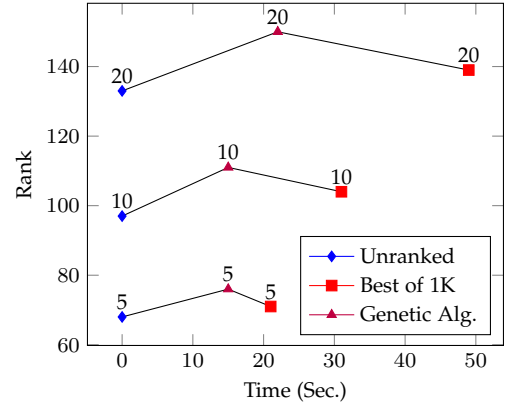

\subsubsection{Bayesian Risk-Reduction Evaluation}
We performed the same analysis as in \autoref{sec:approach:generalized} and similar to \autoref{fig:max_variance}, the variance converges to zero as expected, indicating that our risk estimation gets very accurate after a reasonable number of tests.

There are a total of 1,728 theoretical coverage criteria, which is equivalent to 12 cubed. However, when examining the Moodle example and accounting for our testing model's fixed order and position of four events, only 512 (or eight cubed) of the $C_i$s intersect with $P$.
We generated $30{,}000$ test cases, which gave us $1{,}568$ unique test cases. We repeated the experiment shown in \autoref{fig:max_variance} and received a similar graph only with fewer tests and the best variance of risk estimator approaching $10^{-5}$.

\section{Related Work}

\subsection{Coverage-Criterion Specification}
Producing all possible paths (e.g.,~\cite{rocha2021model}) does not seem to be practical in large systems. Thus, one of the critical issues is the definition of the equivalence classes of tests according to the coverage criterion.

L{\"u}bke et al.~\cite{lubke2019effectiveness} defined equivalence classes by dividing the tested system parameters into two types --- parameters with a limited, small number of options (e.g., a list of payment methods) and parameters with many options (e.g., a bank account number or first name).
It is necessary for this approach to test each parameter of the first type and at least one of the second type. Thus, all models of the second type are in the same equivalence class. 

In some cases, such as autonomous driving, coverage requirements are enforced by regulation. 
To address such large-scale systems,~\cite{li2021comopt} realized the optimization of the testing process based on CTD. Nevertheless, they were still required to reduce the number of overall test scenarios. They achieved that by value quantization and by preventing illogical scenarios. For example, they defined medium traffic load in the range of three to six vehicles and detected illegal scenarios like making a left turn on a straight road.

\subsection{Test-Suite Generation} 
Kuhn and Higdon proposed the first sequence-testing variant, known as $t$-way sequence testing, that allows only one event triggering in a test case~\cite{kuhn2010practical}. Later versions allowed multiple triggered events and extended the algorithms to support additional features~\cite{sheng2018extended, duan2019approach, bombarda2020automata}. Current sequence testing methods for valid test suite generation consist of two steps. The first is to generate a list of all relevant sequences of a length $t$ (called ``target sequences''), and the second step is to generate test cases that cover the target sequences (called ``test sequences'')~\cite{kuhn2010practical}. The latter uses a greedy algorithm that handles constraints between two events. Then, a labeled transition system is proposed to model the SUTs' requirements, and graph path methods are used to find the optimal valid test cases. Based on this work, additional work has been done to expand the language of constraints by, e.g., adding the possibility of contiguous values~\cite{sheng2018extended} and allowing more complex relationships between more than two factors~\cite{duan2019approach}. 

Two problems remain open with these sequence-testing approaches. First, these approaches rely on the existing test model (not part of their work). Second, the solution must be effective in run time and size. To address these challenges, Bombarda and Gargantini~\cite{bombarda2020automata} proposed to model the SUT by a finite state machine and to create the test cases using automata theory. 

Some of the aforementioned papers have shown that generating test cases to cover every possible combination of input parameters is not necessary. This is due to the principle of t-way testing, which seeks to identify the combinations of input parameters that could result in failures. By selecting a small value of t, which represents the number of parameters in each combination, a much smaller test set can be generated compared to exhaustive testing.

\subsection{Utilizing knowledge from previous runs}
Our approach randomly generates test sequences in a black-box manner (we used white-box events to verify the test result in \autoref{sec:evaluation}, however, the tests were generated in a black-box manner). Random black-box test generators, or fuzzes, as they are sometimes called, have evolved to include a feedback loop that facilitates the efficient discovery of bugs, reduction of risk, and increase of coverage. The feedback loop may consist of information on newly achieved code-coverage objectives or the occurrences of desired events during execution, such as buffer overflow~\cite{Candea2019}.  
 
Other ideas for directing the test generation include: choosing inputs far away from the previous inputs~\cite{ARTfar}, dividing the inputs into sub-domains, and using translation to obtain a new test in a different sub-domain~\cite{ChenTsongYueh2007Otcd}. Another approach defines exclusion zones around tests that were already executed and discard randomly chosen inputs if they are chosen from an excluded zone~\cite{Ahmad2014NewSF}. 
 
Other approaches attempt to generate inputs that are more likely to cause a failure based on failure models. For example, boundary inputs are preferred as it is well known that they are more likely to cause a failure~\cite{Ahmad2014NewSF}. Another example chooses test sequences that mutate the values of fields in the object under test~\cite{DBLP:conf/kbse/ZhengZLX10}.


%

\section{Conclusion}
In conclusion, recall the famous quote: ``I always thought something was fundamentally wrong with the universe.'' (The Hitchhiker's Guide to the Galaxy series~\cite{adams1981the}). This feeling is familiar to everyone in the software testing industry. In this paper, we addressed the dilemma of every testing team: what and how much to test before we declare our software as ready to launch?

To tackle this dilemma, we identified three related issues and presented theoretical and practical contributions:
\begin{enumerate}
\item \textbf{How to specify the test space that needs to be covered:} We defined a generalized, automata-based approach for specifying coverage criteria. We also provided a set of valuable coverage criteria that may be applied to various domains.

\item \textbf{Finding a finite, relatively small, test suite that covers this space:} 
We developed a tool that allows us to translate the coverage criteria as ranking functions, generate test suites for these functions, and analyze the results from various aspects. 

\item \textbf{How to utilize knowledge from previous runs to optimally reduce bug risks:} We proposed a Bayesian-based formula for balancing exploration and exploitation of the knowledge obtained by tests.
\end{enumerate}

The approach we presented relies on the existence of a system model. We used the behavioral-programming paradigm to create this model, though other approaches can be used. Evaluating the model construction effort and the paradigm's applicability for modeling large-scale systems, are out of the scope of this paper.
Nevertheless, as demonstrated in~\autoref{sec:eval:moodle}, three small and intuitive b-threads generate a large state space and are sufficient for detecting bugs. This is due to the ability to specify behavioral aspects using small b-threads, without explicitly specifying the cohesive behavior. Furthermore, our approach does not require state-space generation since BP models are executable. Finally, the agility of the paradigm allows for incrementally adding more testing requirements, without changing the existing model.

From what we have presented in this article, it is possible to expand to other research areas in different directions in the testing space. All of these are focused on achieving the goal of efficiently testing processes, software, and system. A possible trend we have begun to explore is the system modeling process concerning the testing resulting from system requirements using BP tools~\cite{weiss2021testing}. Another way to expand is by using statistical tools in the stem of an efficient and focused test suites generation process to test coverage, increasing the probability of identifying the faults.

\ifCLASSOPTIONcompsoc
  \section*{Acknowledgments}
\else
  \section*{Acknowledgment}
\fi
This research was partially supported by grant \#2714/19 from the Israeli Science Foundation. 

The authors would like to thank Gal Amram for a thorough review of the draft of this paper. 


\IEEEtriggeratref{19}

\bibliographystyle{IEEEtranS}
\bibliography{references,achiya}
\end{document}